# Quantum Logic for Trapped Atoms via Molecular Hyperfine Interactions

Gavin K. Brennen and Ivan H. Deutsch

*Center for Advanced Studies, Department of Physics and Astronomy,*

*University of New Mexico, Albuquerque, New Mexico 87131*

Carl J. Williams

*Atomic Physics Division,*

*National Institute of Standards and Technology, Gaithersburg, Maryland 20899*

**Abstract**

We study the deterministic entanglement of a pair of neutral atoms trapped in an optical lattice by coupling to excited-state molecular hyperfine potentials. Information can be encoded in the ground-state hyperfine levels and processed by bringing atoms together pair-wise to perform quantum logical operations through induced electric dipole-dipole interactions. The possibility of executing both diagonal and exchange type entangling gates is demonstrated for two three-level atoms and a figure of merit is derived for the fidelity of entanglement. The fidelity for executing a CPHASE gate is calculated for two $^{87}$Rb atoms, including hyperfine structure and finite atomic localization. The main source of decoherence is spontaneous emission, which can be minimized for interaction times fast compared to the scattering rate and for sufficiently separated atomic wavepackets. Additionally, coherent couplings to states outside the logical basis can be constrained by the state dependent trapping potential.

# I. Introduction

Over the last few decades tremendous progress has been made in coherent control and manipulation of individual quantum systems in atomic, molecular, and optical (AMO) physics. Motivated primarily by the goal of improving precision measurement and noise reduction, as well as testing the foundations of quantum theory, quantum opticians have developed a variety of new methods and systems [1], including laser cooling of atoms, ion traps, optical lattices, cavity QED, atom interferometers, and correlated photon sources. Simultaneously, physical chemists have been perfecting techniques for coherent control of molecular reactions and other complex systems via ultra-fast laser spectroscopy [2]. Today, these varieties of tools are converging on a new problem – quantum information processing (QIP) [3,4]. The ability to coherently control a many-body system has great potential for new paradigms in computation, communication, and precision measurement. The unique properties of AMO physical systems make them the ideal arena for implementing these ideas.

One particularly attractive system in this context is laser cooled and trapped neutral atoms. Of crucial importance is the ability to perform deterministic entanglement via two-atom interactions (e. g. a controlled-NOT quantum logic gate). This has been discussed for several different dynamical interactions such as ground-state collisions of atoms [5,6], and induced electric dipole-dipole interactions [7,8], including highly excited Rydberg states [9]. The common goal of these proposals is to design a protocol with a flexible trapping architecture, a means to encode quantum information in the atoms, an ability to carry out quantum logic via atomic interaction with minimal loss of information to the decohering environment, and a faithful read-out protocol.

While the impetus for much of this research has been the pursuit of multiparticle entanglement for QIP, the search for such encodings and two body interactions yields insight into new areas of research that unite ideas of atomic, molecular physics, and coherent chemistry. An example of research in this area is "superchemistry", where coherent coupling between two separated atoms and molecular dimer states has been observed in a BEC [10]. The goal of coherent control of a molecular dimer can be put in one-to-one correspondence with the problem of implementing two-qubit quantum logic gates, as we will show below.

We study here alkali atoms in tight traps that interact pair-wise by induced dipole-dipole interactions in a far off-resonance bulk 3D optical lattice. Various other trapping schemes such as magnetic [11] or optical microtraps [12] might be used as such technologies mature. As described in [8], a suitable geometry consists of independent linearly polarized standing waves, of slightly different frequencies along the three Cartesian axes where atoms are trapped at the nodes of blue detuned standing waves. Along a defined $\hat{z}$-axis (quantization axis), one can vary the relative angle θ between polarization vectors of the counterpropagating beams, and the field decomposes into $\sigma_\pm$ standing waves whose nodes are separated by $\lambda(\theta/2\pi)$. We identify two "species" of atoms denoted (±), that are trapped in predominately $\sigma_\pm$ light. A logical basis for each species is defined,

$$|0\rangle_\pm = |S_{1/2},(F_\downarrow, m_F = \mp 1)\rangle \otimes |\psi_\pm\rangle_{ext}, \quad |1\rangle_\pm = |S_{1/2},(F_\uparrow, m_F = \pm 1)\rangle \otimes |\psi_\pm\rangle_{ext}, \qquad (1)$$

where $|S_{1/2},(F,m_F)\rangle$ is a particular magnetic sublevel of the ground hyperfine manifold (with $F_{\uparrow,\downarrow} = I \pm 1/2$), $I \geq 3/2$ is the nuclear spin, and $|\psi_\pm\rangle_{ext}$ is the external coordinate wavefunction for the (±) species [13]. For simplicity we assume that each of the atoms is prepared in the ground motional state of a locally isotropic trapping potential. As the laser polarization angle is varied from θ=π/2 to near 0°, atoms prepared in these logical basis states will adiabatically follow the moving $\sigma_\pm$ standing waves and be brought together pair-wise. When the atoms are sufficiently close to one another, one can apply an external pulse – which we refer to as the "catalysis" field – inducing electric dipoles in the two atoms that are stronger than those induced by the trapping field, and causing the atoms to evolve in a nonseparable manner. After the desired interaction time, the catalysis laser is turned off and the atoms are separated again. If the coherent interaction is strong enough, then the time to perform the entangling gate can be much shorter than incoherent processes such as photon scattering and inelastic two-body collisions. Under these circumstances the gate can be executed with high fidelity.

This discussion assumes the individual atoms maintain their identical structure during the interaction. At small internuclear distances where the highest fidelity for two-qubit operations occurs, a proper characterization of the interaction of the catalysis with the two-atom system requires us to consider the *molecular* spectrum. A molecular treatment has considerable complexity, especially when hyperfine interactions are included in the description, but is essential when we encode in terms of these quantum numbers. Our goal here is to calculate the molecular potentials and oscillator strengths of states that asymptotically connect to $S_{1/2}(F) + P_{1/2}(F')$ atoms (we consider here $^{87}$Rb, with $I=3/2$). We begin in Sec. II by presenting a simplified model of the dipole-dipole interactions for three level atoms. This elucidates many of the important properties of the more detailed and complete model presented in Sec. III. The results characterizing the regime of optimal fidelity for producing deterministic entanglement are discussed in Sec. IV and a summary is given in Sec. V.

## II. SIMPLIFIED MODEL: THREE-LEVEL ATOMS

The essence of our system is to encode information in the ground electronic hyperfine states and induce interaction between the atoms by mixing in (via the catalysis pulse) a small amplitude of excited electronic states. The simplest model which contains these elements consists of two atoms (labeled $\alpha$ and $\beta$) each with 3-levels: a "ground-state" doublet basis $|g\rangle = \{|0\rangle, |1\rangle\}$, split by an energy $\hbar\omega_{01}$, and an "excited-state" $|e\rangle$ with an "optical" energy difference $E_e - E_0 = \hbar\omega_{0e}$, as illustrated in Fig. 1a. After tracing over the vacuum modes in the Born-Markov approximation, the effective non-Hermitian Hamiltonian is [14], $H = H_A + H_{AL} + H_{dd}$, consisting of the bare atomic Hamiltonian for a pair of non-interacting atoms, atom-laser interaction and dipole-dipole coupling. Taking the zero of energy at $|0\rangle$, the first two terms are,

$$H_A = \left[\hbar\omega_{01}|1\rangle\langle 1| - \hbar(\Delta + i\Gamma/2)(c_0^2 + c_1^2)|e\rangle\langle e|\right]_\alpha \otimes \mathbf{1}_\beta$$
$$+ \mathbf{1}_\alpha \otimes \left[\hbar\omega_{01}|1\rangle\langle 1| - \hbar(\Delta + i\Gamma/2)(c_0^2 + c_1^2)|e\rangle\langle e|\right]_\beta \quad (2)$$

$$H_{AL} = -\frac{\hbar\Omega}{2}\sum_{g=0}^{1}\left(D_{\alpha g}^\dagger + D_{\beta g}^\dagger\right) + h.c.,$$

where $\Delta = \omega_c - \omega_{e0}$ is the catalysis laser detuning, $\Gamma$ and $\Omega$ are the excited-state decay rate and Rabi frequency with unit oscillator strength, and $D^\dagger_{vg} = c_g(|e\rangle\langle g|)_v$ is the dimensionless dipole raising operator for atom $v=\alpha,\beta$ connecting ground-state g=0,1 to the excited-state with oscillator strength $c_g$ (taken to be real). The dipole-dipole coupling Hamiltonian is

$$H_{dd} = \left(V_c - i\frac{\hbar\Gamma_c}{2}\right)\sum_{g,g'=0}^{1}\left(D_{\alpha g}^\dagger D_{\beta g'} + D_{\beta g'}^\dagger D_{\alpha g}\right), \quad (3)$$

where $V_c$ is the coupling strength which depends implicitly on $r$, and $\Gamma_c$ is the collective contribution to the decay rate; *i.e.* the degree to which the molecular decay rate is modified from that of a free atom.

Partial diagonalization of $H_A + H_{dd}$, for $V_c \ll \hbar\omega_{01}$ yields "molecular eigenstates"

$$\left\{|00\rangle, |01\rangle, |10\rangle, |11\rangle, |ge\rangle_\pm \equiv \frac{|ge\rangle \pm |eg\rangle}{\sqrt{2}}, |ee\rangle\right\}, \quad (4)$$

where g={0,1}. In this basis, the Hamiltonian is $H = H_0 + H_{AL}$ with

$$H_0 = \hbar\omega_{01}(|01\rangle\langle01|+|10\rangle\langle10|) + 2\hbar\omega_{01}|11\rangle\langle11| - 2(\hbar\Delta + i\hbar\Gamma/2)|ee\rangle\langle ee| +$$
$$(\pm c_0^2 V_c - i\hbar\Gamma_{0\pm}/2 - \hbar\Delta)|0e\rangle_{\pm\pm}\langle 0e| + (\pm c_1^2 V_c - i\hbar\Gamma_{1\pm}/2 - \hbar\Delta + \hbar\omega_{01})|1e\rangle_{\pm\pm}\langle 1e| +$$
$$\pm c_0 c_1 (V_c - i\hbar\Gamma_c/2)(|0e\rangle_{\pm\pm}\langle 1e| + |1e\rangle_{\pm\pm}\langle 0e|)$$

$$H_{AL} = -\frac{\hbar\Omega}{\sqrt{2}}\left[c_0\left(|0e\rangle_+\langle 00| + |ee\rangle_+\langle 0e| - \frac{1}{2}(|1e\rangle_-\langle 01| - |1e\rangle_-\langle 10|) + \frac{1}{2}(|1e\rangle_+\langle 01| + |1e\rangle_+\langle 10|)\right) \right.$$
$$\left. + c_1\left(|1e\rangle_+\langle 11| + |ee\rangle_+\langle 1e| + \frac{1}{2}(|0e\rangle_-\langle 01| - |0e\rangle_-\langle 10|) + \frac{1}{2}(|0e\rangle_+\langle 01| + |0e\rangle_+\langle 10|)\right) + h.c.\right].$$

(5)

The symmetric states $|ge\rangle_+$ are superradiant with linewidths $\Gamma_{g+} = \Gamma + c_g^2 \Gamma_c$, and couple to $|gg\rangle$ and $|ee\rangle$ with Rabi frequency $c_g\sqrt{2}\Omega$. The states $|ge\rangle_-$ are subradiant with linewidths $\Gamma_{g-} = \Gamma - c_g^2 \Gamma_c$. In the case of two 2-level atoms, the subradiant state is *dark* to the atom-laser interaction. For multilevel atoms, however, super and sub-radiant states in Eq. (2), which are asymptotically split by the ground-state energy, are no longer eigenstates of $H_A + H_{dd}$. Rather, they mix under the dipole-dipole interaction, and for $V_c \sim \hbar\omega_{01}$, this mixing allows the degenerate ground-states $|01\rangle$, $|10\rangle$ to effectively interact. In the far detuned or weak field limit the effects of the doubly excited $|ee\rangle$ can be ignored.

We consider level shifts induced on the ground-states through adiabatic elimination of the excited-states, valid under the conditions of low saturation. The reduced non-Hermitian Hamiltonian $H'$ is found for the dressed ground-state subspace to first order in $V_c/\hbar\omega_{01}$,

$$H'_{00,00} = 2c_0^2 \Lambda(\delta_{20}, \Gamma_{0+})$$
$$H'_{01,01} = c_0^2 \Lambda(\delta_{11}, \Gamma_{1-})/2 + c_0^2 \Lambda(\delta_{21}, \Gamma_{1+})/2 + c_1^2 \Lambda(\delta_{30}, \Gamma_{0-})/2 + c_1^2 \Lambda(\delta_{40}, \Gamma_{0+})/2$$
$$H'_{01,10} = c_0^2 \Lambda(\delta_{21}, \Gamma_{1+})/2 - c_0^2 \Lambda(\delta_{11}, \Gamma_{1-})/2 + c_1^2 \Lambda(\delta_{40}, \Gamma_{0+})/2 - c_1^2 \Lambda(\delta_{30}, \Gamma_{0-})/2 \quad (6)$$
$$H'_{10,10} = H'_{01,01}, \quad H'_{10,01} = H'_{01,10}$$
$$H'_{11,11} = 2c_1^2 \Lambda(\delta_{41}, \Gamma_{1+}),$$

where the complex energy scale of the perturbation is,

$$\Lambda(\delta,\Gamma) = \frac{\hbar|\Omega|^2}{4(\delta + i\Gamma/2)}. \tag{7a}$$

with "molecular" detunings,

$$\delta_{1g} = c_g^2 V_c/\hbar + \Delta, \quad \delta_{2g} = -c_g^2 V_c/\hbar + \Delta, \quad \delta_{3g} = c_g^2 V_c/\hbar + \Delta + \omega_{01}, \quad \delta_{4g} = -c_g^2 V_c/\hbar + \Delta + \omega_{01} \tag{7b}$$

(see Fig. 1b). In the limit $V_c \to 0$ at infinite interatomic separation, the exchange coupling $H'_{01,10}$ vanishes, and the reduced Hamiltonian is separable, as expected. In this case, we recognize the real part of $\Lambda$ to be the atomic light shift and the imaginary part the photon scattering rate. The nonseparable interaction at finite interatomic separation leads to entanglement.

The dressed Hamiltonian $H'$ can be used to create deterministic entanglement within the internal states of the two atoms via the exchange interaction $H'_{01,10}$. This is the case studied in [15] for the dipole-dipole interaction between atoms with zero nuclear spin and degenerate ground-states $|S_{1/2}, m_S = \pm 1/2\rangle$.

The dipole-dipole interaction $H'$ can also produce entanglement without swapping the states of the constituent atoms. As we will discuss below, for real alkali atoms trapped in an optical lattice the entanglement based swapping can be strongly suppressed because of imperfect spatial wave function overlap for these transitions. In this case the interaction is approximately diagonal and the universal CPHASE [4] can be implemented by allowing the induced dipoles to interact for a time $\tau = \hbar\pi/|\text{Re}[E_{00} + E_{11} - 2E_{01}]|$, where $E_{ij} = \langle i,j|H'|i,j\rangle$ are the complex diagonal matrix elements. Note, for a separable interaction $E_{ij} = E_i + E_j$, and thus the required gate time goes to infinity as expected.

The probability that the desired entangling gate was successfully performed can be measured by the fidelity $\mathcal{F} = \min_{|i\rangle} |\langle i|U^\dagger V_{\text{eff}}|i\rangle|^2$ where $U$ is the desired unitary transformation (here

CPHASE), $V_{eff}$ is the nonunitary operator generated by the effective Hamiltonian including decay for the interaction, and the minimum is taken over all possible input states. For large enough atomic separations the dominant source of decoherence is from spontaneous emission which occurs from each state at a rate $\gamma_{ij} = 2\,\text{Im}[E_{ij}]$. The fidelity for the CPHASE gate in the worst case scenario is

$$\mathcal{F}_{\text{CPHASE}} = \exp\left(-\left(\gamma_{ij}\right)_{\max}\tau\right) = \exp\left(-\frac{2\pi\,\text{Max}\left(\text{Im}[E_{ij}]\right)}{\left|\text{Re}[E_{00}] + \text{Re}[E_{11}] - 2\,\text{Re}[E_{01}]\right|}\right) \equiv e^{-1/\kappa}, \qquad (8)$$

where the figure of merit $\kappa$ is the ratio of the coherent levels shifts to the spontaneous linewidth as described in our previous analyses [8].

We can analytically express exactly the behavior of $\kappa$ versus the parameters of the two-atom problem, but the results are more transparent under certain approximations. Specifically, given a ground-state splitting small compared to the laser detuning but large compared to the dipole-dipole coupling, $|\Delta| \gg \omega_{01} \gg |V_c|/\hbar$, the figure of merit to first order in $\omega_{01}/\Delta$ is

$$\kappa \approx \left|\frac{V_c\left(c_0^2 - c_1^2\right)}{\hbar\pi}\,\text{Min}\left[\frac{c_0^2 + c_1^2(2\omega_{01}/\Delta - 1)}{c_0^2\left(c_0^2 + 1\right)}, \frac{c_0^2(2\omega_{01}/\Delta + 1) - c_1^2}{c_1^2\left(c_1^2 + 1\right)}\right]\right|. \qquad (9)$$

Operations to achieve highest fidelity depend on the details of this model. It is evident that the figure of merit is very sensitive to the relative oscillator strengths and contains a term that scales inversely with the detuning of the catalysis laser from free atomic resonance. Thus the performance of the gate depends both on geometry, through the interatomic separation $r$ since $V_c \sim 1/r^3$, and the strength of the induced dipole moments.

8CPHASE), $V_{eff}$ is the nonunitary operator generated by the effective Hamiltonian including decay for the interaction, and the minimum is taken over all possible input states. For large enough atomic separations the dominant source of decoherence is from spontaneous emission which occurs from each state at a rate $\gamma_{ij} = 2\,\text{Im}[E_{ij}]$. The fidelity for the CPHASE gate in the worst case scenario is

$$\mathcal{F}_{\text{CPHASE}} = \exp\left(-\left(\gamma_{ij}\right)_{\max}\tau\right) = \exp\left(-\frac{2\pi\,\text{Max}\left(\text{Im}[E_{ij}]\right)}{\left|\text{Re}[E_{00}] + \text{Re}[E_{11}] - 2\,\text{Re}[E_{01}]\right|}\right) \equiv e^{-1/\kappa}, \qquad (8)$$

where the figure of merit $\kappa$ is the ratio of the coherent levels shifts to the spontaneous linewidth as described in our previous analyses [8].

We can analytically express exactly the behavior of $\kappa$ versus the parameters of the two-atom problem, but the results are more transparent under certain approximations. Specifically, given a ground-state splitting small compared to the laser detuning but large compared to the dipole-dipole coupling, $|\Delta| \gg \omega_{01} \gg |V_c|/\hbar$, the figure of merit to first order in $\omega_{01}/\Delta$ is

$$\kappa \approx \left|\frac{V_c\left(c_0^2 - c_1^2\right)}{\hbar\pi}\,\text{Min}\left[\frac{c_0^2 + c_1^2(2\omega_{01}/\Delta - 1)}{c_0^2\left(c_0^2 + 1\right)}, \frac{c_0^2(2\omega_{01}/\Delta + 1) - c_1^2}{c_1^2\left(c_1^2 + 1\right)}\right]\right|. \qquad (9)$$

Operations to achieve highest fidelity depend on the details of this model. It is evident that the figure of merit is very sensitive to the relative oscillator strengths and contains a term that scales inversely with the detuning of the catalysis laser from free atomic resonance. Thus the performance of the gate depends both on geometry, through the interatomic separation $r$ since $V_c \sim 1/r^3$, and the strength of the induced dipole moments.

The simplified model described in this section highlights many important features of the dipole-dipole interaction between real alkalis. Specifically, we find that under the adiabatic approximation the interaction allows couplings which can change internal ground-states, or if these exchanges are suppressed, it can produce entanglement through a diagonal interaction acting on the logical basis states that induces differential level-shifts. This flexibility is an advantage when one wants a two-qubit gate with high fidelity. There are several limitations to this model, however, that require the inclusion of the internal structure of alkali atoms.

First, the asymptotic argument yielding Eq. (9), describes the behavior of fidelity for *weak* dipole-dipole interactions. However, we will see below that the region of best fidelity for trapped alkalis with hyperfine structure is $V_c \sim \hbar\omega_{01}$. Second, the above model treats the atoms as point particles, when in reality they are localized wave packets with finite extent set by the trapping potentials. Thus, there is always a finite probability for atoms to be separated by a "Condon radius" – the internuclear separation at which the catalysis is on resonance with one of the molecular potentials. The "Condon radius" can be viewed as an intermolecular dependent detuning that can lead to enhanced spontaneous emission resulting from resonant molecular excitation. The design of the entangling gate must balance the need to bring the atoms close together in order to obtain a large dipole-dipole interaction, while simultaneously maintaining sufficient separation so that there is negligible probability to be at a Condon radius. Finally, the three-level model treats $V_c$ as a scalar when in fact the dipole-dipole interaction depends on the orientation of the induced dipoles relative to the internuclear separation. In order to take these important features into account, a more complete calculation is required, as discussed in the next section.

### III. MOLECULAR HYPERFINE STRUCTURE

An appropriate set of "good" quantum numbers for describing the molecular potentials depends on the strengths of the atom-atom interaction as a function of internuclear separation compared to the intra-atomic energy scales (e.g. optical S-P transitions, and fine or hyperfine interactions). Our description will be determined by a choice that gives the best fidelity for performing quantum logic with the information encoding according to Eq. (1). In order to maintain the logical basis, we must preserve the ground-state hyperfine quantum numbers. In our system, all interactions are mediated through virtual transitions to the excited *S+P* manifold whose energy levels are shifted by the dipole-dipole interaction. We thus require that the dipole-dipole shift never be much greater than the ground-state hyperfine splitting at the distances spanned by the relative coordinate probability distribution. At these separations the *excited-state* hyperfine structure of alkali atoms is small relative to the dipole-dipole interaction and therefore the excited state hyperfine labels no longer represent good quantum numbers. We operate here at relatively large internuclear separations beyond the Hund's case (c) [16] conditions, where dipole-dipole shifts are small compared to spin-orbit coupling and large compared to hyperfine shifts. Because the dipole-dipole interaction induces mixing among the atomic orbitals, an atomic product basis set describing a given fine structure asymptote is inappropriate.

We restrict our attention then to the molecular potentials that asymptotically connect to the multiplet of hyperfine levels associated a given fine-structure manifold. For simplicity we consider the D1 line in alkalis, $S_{1/2} + P_{1/2}$. The Born-Oppenheimer Hamiltonian for these states can then be expressed as

$$H_{S_{1/2}+P_{1/2}} = H_{S_{1/2}} + H_{P_{1/2}} + H_{hf} + H_{dd}, \qquad (10)$$

where $H_{S_{1/2}}, H_{P_{1/2}}, H_{hf}$ describe the energy levels of the atomic orbitals, including the hyperfine interaction, and $H_{dd} = V_{dd} - i\hbar\Gamma_{dd}/2$ is the dipole-dipole coupling in the near field,

$$V_{dd} = \sum_q \frac{(-1)^q d^\dagger_{\alpha\,q} d_{\beta\,-q} - 3 d^\dagger_{\alpha\,0} d_{\beta\,0}}{r^3} + h.c., \quad \Gamma_{dd} = \frac{\Gamma}{2}\sum_q (-1)^q \left(D^\dagger_{\alpha\,q} D_{\beta\,-q} + D^\dagger_{\beta\,q} D_{\alpha\,-q}\right), \qquad (11)$$

where the $d_v$ are the actual electric dipole operators (with dimensions) for each atom. Here the dipole operators are described with a quantization axis along the *internuclear* (body fixed) axis. Diagonalizing as a function of *r* yields the Born-Oppenheimer molecular potentials. In principle, Eq. (10) should also include rotational energy of the dimer, $H_{rot} = \hat{l}^2/(2mr^2)$. Each partial wave component in the ground-state will couple to the appropriate rotational states in the excited-state. We ignore this effect for two reasons. First, we consider separated atoms such that $H_{rot} \ll H_{hf}$ over the range of probable internuclear separations. Thus, the manifold of rotational levels can be treated as nearly degenerate. Second, we consider *trapped atoms* prepared in the vibrational ground-state and we assume the light shift induced by the catalysis field to be a perturbation to the optical lattice. Rotations of the dimer would correspond to coherent couplings to higher vibrational levels in the ground-state, via mixing with the excited-states. These are suppressed by an energy gap equal to the trap oscillator energy. In other words, any ground-state wavepacket reshaping by adiabatic mixing with the untrapped motional states in the excited-states is suppressed by the trapping potential. In this way the couplings to higher rotational states in the excited-state manifold are effectively calculated as an incoherent sum over degenerate eigenstates and can only act as an additional weak internuclear dependent shift.

To find the molecular potentials and eigenstates, we start with the asymptotic ($r \to \infty$) basis of eigenstates. These are symmetric and antisymmetric states with respect to exchange of the two atomic orbitals, denoted with quantum number $\pi = \pm 1$,

$$\left|S_{1/2}(F,m_F),P_{1/2}(F'm_{F'});\pi\right\rangle = \frac{1}{\sqrt{2}}\left(\left|S_{1/2}(F,m_F)\right\rangle_\alpha \left|P_{1/2}(F',m_{F'})\right\rangle_\beta + \pi\left|P_{1/2}(F',m_{F'})\right\rangle_\alpha \left|S_{1/2}(F,m_F)\right\rangle_\beta\right),$$

(12)

with all magnetic quantum numbers defined with respect to the internuclear axis. In this basis the dipole-dipole interaction has the matrix representation,

$$\langle S_{1/2}(F_j m_{Fj}) P_{1/2}(F'_j m_{F'j}) \pi | V_{dd} | S_{1/2}(F_i m_{Fi}), P_{1/2}(F'_i m_{F'i}); \pi \rangle = \pi \frac{2d^2}{r^3} A \quad (13a)$$

where $d$ is the reduced matrix element of the atomic dipole operator, and the indices $i$ and $j$ label the quantum numbers for the initial and final states. The coefficient $A$ accounts for the angular momentum coupling for this tensor operator,

$$A = (-1)^{F_i+F_j} \sqrt{(2F_i+1)(2F_j+1)} \begin{Bmatrix} F_i & I & 1/2 \\ 1/2 & 1 & F_j \end{Bmatrix} \begin{Bmatrix} F'_j & I & 1/2 \\ 1/2 & 1 & F_i \end{Bmatrix}$$
$$\times \sum_q c^{F_j,1,F'_i}_{m_{Fj},q,m'_{F'i}} c^{F_i,1,F'_j}_{m_{Fi},q,m'_{F'j}} - 3 c^{F_j,1,F'_i}_{m_{Fj},0,m_{Fi}} c^{F_i,1,F'_j}_{m_{Fi},0,m_{Fj}}, \quad (13b)$$

where the $c$'s are Clebsch Gordan coefficients $c^{F,1,F'}_{m,q,m'} = \langle Fm,1q|F'm' \rangle$, and the terms in curly brackets are Wigner 6 j-symbols. Ignoring rotational effects, the interaction obeys the selection rule

$$m_{Fi} + m_{F'i} = m_{Fj} + m_{F'j} \equiv M_{tot} \quad (14)$$

corresponding to conservation of total magnetic projection along the internuclear axis. This is required by Eq. (11), where $V_{dd}$ is proportional to $1/r^3$ times the second rank spherical harmonic $Y_2^0$. Further, $V_{dd}$ is invariant under a change of the sign of $M_{tot}$, amounting to invariance under interchange of the two atoms though the diatomic origin. The excited-state eigenvalues and eigenvectors are calculated by diagonalizing Eq. (13) in blocks labeled by $M_{tot}$ and $\pi$. Note $M_{tot}$ is not conserved in the situation where rotational excitation of the atomic fragments is not suppressed by the trapping potential.

We consider two $^{87}$Rb atoms ($I = 3/2$) driven by a catalysis laser detuned from the D1 resonance $5S_{1/2} \to 5P_{1/2}$. In the $S_{1/2} + P_{1/2}$ manifold, including hyperfine interactions with energy splitting $V_{hf}(S_{1/2}) = 1263.4\,\hbar\Gamma$, $V_{hf}(P_{1/2}) = 151.2\,\hbar\Gamma$ ($\Gamma = 2\pi \times 5.41$MHz), there are 128 properly symmetrized atomic basis states. The resulting 128 molecular potentials are plotted in Fig. 2, and clearly correlate to the four asymptotic combinations of atomic hyperfine energy levels as $r \to \infty$, and to six Hund's case (c) states for $kr < 0.05$.

For weak saturation, we treat the dipole-dipole interaction as a perturbation to the trapping potential, and the excited-state molecular potentials can be adiabatically eliminated. Given a coupling strength defined by atomic Rabi frequency $\Omega$, the reduced "dressed" Hamiltonian in the ground-state basis $(i,j)$ is

$$H_{ij} = \frac{\hbar|\Omega|^2}{4} \left\langle \sum_{|e(\mathbf{r})\rangle} \frac{c^*_{ei}(\mathbf{r}) c_{ej}(\mathbf{r})}{\delta_e(r) + i\gamma_e(r)} \right\rangle_{rel}. \tag{15}$$

The sum is taken over all $|e(\mathbf{r})\rangle$, the Born-Oppenheimer internal "molecular" states at $\mathbf{r} = r\hat{z}_{BF}$ where $\hat{z}_{BF}$ is the body-fixed internuclear axis. The position dependent molecular oscillator strengths, detunings, and decay rates are defined,

$$c_{ie}(\mathbf{r}) = \langle e(\mathbf{r}) | \vec{D}^\dagger \cdot \vec{\varepsilon}_c | i \rangle, \quad \delta_e(r) = \Delta - \lambda_e(r), \quad \gamma_e(r) = \Gamma/2 + \langle e(\mathbf{r}) | \Gamma_{dd} | e(\mathbf{r}) \rangle. \tag{16}$$

Here, $\vec{\varepsilon}_c$ is the space-fixed catalysis polarization, the "atomic" detuning is defined with respect to the $|S_{1/2}, F=1\rangle \to |P_{1/2}, F'=1\rangle$ resonance, $\lambda_e(r)$ are the Born-Oppenheimer eigenvalues of Eq.

(10) relative to the $S_{1/2} + P_{1/2}$ asymptote, and $\Gamma_{dd}$ is the near field cooperative part of the decay defined in Eq. (11).

The average in Eq. (15) is taken over the relative coordinate probability distribution of the atomic pair. To calculate this expression, it is necessary to integrate over all relative orientations of the interatomic separation **r** with respect to the *space-fixed* (SF) axis $\hat{z}$. Assuming a catalysis beam $\pi$-polarized with respect to the SF axis, the atomic ground-states $|S_{1/2};(F,\mu_F)\rangle_{SF}$ will couple to excited-states $|P_{1/2};(F',\mu_{F'})\rangle_{SF}$, where we have used $\mu$ to denote the magnetic quantum number with respect to the SF-axis. The molecular eigenstates are calculated as linear combinations of product states quantized along the *body-fixed* (BF) axis. To calculate the expectation with the external coordinate wavefunction we perform a rotation or frame transformation [17] on the excited molecular eigenstates to a SF basis with identical structure,

$$
\begin{aligned}
&|S_{1/2}(Fm_F), P_{1/2}(F'm_{F'});\pi\rangle_{BF} = \\
&\sum_{\mu_F,\mu_{F'}} D^{(F)}_{m_F,\mu_F}(\phi,\theta,0) D^{(F')}_{m_{F'},\mu_{F'}}(\phi,\theta,0) |S_{1/2}(F\mu_F), P_{1/2}(F'\mu_{F'});\pi\rangle_{SF},
\end{aligned}
\qquad (17)
$$

where the arguments of the Wigner rotations matrices, $(\theta,\phi)$, are polar angles between the internuclear coordinate **r** and the space fixed axis $\hat{z}$. Under this transformation, $\vec{\varepsilon}_c \to \vec{\varepsilon}_c(\mathbf{r})$ with components now defined relative to $\hat{z}_{BF}$. The integration then involves the product of a Gaussian for the relative coordinate of the separated atoms with polynomials of trigonometric functions, and can be carried out analytically.

## IV. RESULTS AND DISCUSSION

Using the results from Sec. III we calculate the fidelity $\mathcal{F}$, defined in Eq. (8), for performing a CPHASE gate using trapped $^{87}$Rb atoms. Figure 3 shows a surface plot of $\mathcal{F}$ as a function of catalysis laser detuning relative to atomic resonance $\Delta$, and the separation between the atomic wavepackets $\Delta z$, with localization parameter $\eta \equiv kz_0 = 0.05$, where $z_0$ is the rms width of the ground-vibrational packet along $\hat{z}$. A comparison with Fig. 2 shows that the region of best fidelity occurs for internuclear separations where $V_{dd} \sim V_{hf}$. We calculated the fidelity for positive detunings only because we treat the potentials coupled to in Eq. (15) as supporting a continuum of states. At negative detunings, potentials that scale like $-1/r^3$ support a finite number of bound states [18]. An interesting question that we do not address here is whether one could reach higher fidelities by red detuning the catalysis between bound states of the excited-state potentials. The affect of spontaneous decay for red detunings relative to blue detunings is described in [19]. One additional complexity with detuning to the red is the high density of bound levels at detunings on the order of $\omega_{01}$, especially for the heavier alkalis.

The behavior of the fidelity depends both on the geometry of the separated atomic wavepackets and the strength of the induced dipoles, and can be inferred from the results in Fig. 3. According to Eq. (8), high fidelity in our protocol requires large differential energy level shifts of the logical basis states arising from the different detunings and oscillator strengths which couple the ground molecular potentials to the excited molecular potentials. The bigger the differential shift, the faster the gate, and the less chance for decoherence resulting from spontaneous emission. Such differential couplings are most prominent at small internuclear separations and large detunings where the dipole-dipole coupling yields large splittings between the excited potentials. Of course the Condon points, the internuclear separations at where the catalysis laser is resonant with one of the molecular potentials, should be avoided. To balance these two effects it is found that optimal fidelity for large detunings occurs at wavepacket separations such that the Condon radii lies $\sim 1-3$ rms widths outside the peak of the relative coordinate wavefunction. Wavepacket separations closer than this are not plotted in Fig. 3 as the atoms experience substantial decay and the adiabatic approximation no longer provides a valid description of the wavefunction. At separations just beyond the optimal region, the range of internuclear radii yielding the largest differential couplings lies in the tails of the relative coordinate wavefunction, and the fidelity drops exponentially as is verified by fitting the fidelity to $e^{-1/\kappa}$ for large $\Delta z$. For even larger separations, the wavepackets look like point dipoles and the figure of merit falls off like $1/\Delta z^3$, as expected. This is verified by fitting the fidelity to $e^{-1/\kappa}$ for large $\Delta z$.

The functional dependence of fidelity with detuning depends in a detailed way on the relative oscillator strengths and the ground-state splitting as presaged in the simple three level model of Sec. II. Some features can be understood in a qualitative manner. At detunings in the range $0 < \Delta < 2000\Gamma$ the fidelity is quite poor, reflecting the fact that the Condon radii for these detunings correspond to very large internuclear separations where the excited state potentials are weakly split when compared to the ground-state splitting. Thus, there is not a substantial differential light-shift accumulated on the logical basis states. The small peak in fidelity at $\Delta \sim 1000\Gamma$ corresponds to detuning between the hyperfine splitting of the asymptotic excited-states. The wavepacket separation needed to avoid photon scattering at this detuning is too large to yield a high fidelity. At larger detunings, $\Delta > 2V_{hf}(S_{1/2})$, the fidelity shows a gradual improvement with detuning. This can be understood from the fact that the largest scattering rate scales like $1/\Delta^2$, decreasing slightly faster than the differences of the coherent light-shifts.

There are several constraints that must be satisfied for the model presented here to be self-consistent. First, the gate-time must be short compared to the time to scatter a photon. Our analysis only accounts for possible scattering from the catalysis and completely neglects spontaneous emission from the optical lattice. We thus require that the atomic saturation parameter for the lattice must be small compared to that of the catalysis. This puts a constraint on the peak intensity and detuning of the lattice and catalysis according to, $\eta^2 I_L / \Delta_L^2 << I_c / \Delta_c^2$, where we use the fact that the lattice is blue detuned, so that atoms are trapped at the nodes of the standing waves where the scattering is suppressed by the Lamb-Dicke factor $\eta^2$. Second, we have assumed throughout that the dipole-dipole shift is a perturbation to the trapping potential.

This is ensured by requiring the gate time to be much larger than the oscillation period of the trap, $\tau \gg 2\pi/\omega_{osc}$. We write the gate time as $\tau = \pi/(\xi\Gamma'_c) = 2\pi I_0/(I_c \xi \Gamma)$, where $\xi \equiv \text{Re}[E_{00} + E_{11} - 2E_{01}]/(\hbar\Gamma'_c)$ is the strength of the differential ground-state level shift in units of the photon scattering rate on atomic resonance, $\Gamma'_c = \Gamma I_c/(2I_0)$ with $I_0$ the saturation intensity. Using the relation $\hbar\omega_{osc} = 2\sqrt{2U_0 E_R/3}$ with $U_0$ the maximum light-shift induced by the lattice [20] and $E_R$ the recoil energy, we obtain the constraint,

$$\eta^2 \left(\frac{\Delta_c}{\Delta_L}\right)^2 \ll \frac{I_c}{I_L} \ll \frac{1}{\xi}\left(\frac{\omega_{osc}}{\Gamma}\right)\left(\frac{I_0}{I_L}\right). \tag{18}$$

For the parameters $E_R = \hbar\Gamma/1500$, $\eta = 0.05$, $\Delta_c = 10^4 \Gamma$, we find that at a well separation $k\Delta z = 0.15$ the fidelity is maximum and $\xi \cong 3.5 \times 10^{-7}$. Under these circumstances, Eq. (18) can be satisfied for the following experimentally achievable parameters, $I_L = 10 I_c = 3.2 \times 10^6 I_0$, $\Delta_L = 10^4 \Gamma$ which would result in a gate speed $1/\tau \cong 0.1(\omega_{osc}/2\pi) = 144\,\text{kHz}$.

In the above calculation of fidelity, imperfect operation arose solely from spontaneous emission of the excited quasimolecule. There are, of course, many other sources that degrade performance, even if one neglects technical error. For example, off-diagonal couplings, both within and outside of the computational basis, correspond to errors in the CPHASE gate. The latter is typically referred to as "leakage". Off-diagonal transitions can be induced by the dipole-dipole interaction or though ground-state scattering. We focus first on the former mechanism and show how it can be suppressed by the geometry of the trapping potential to a degree that affords sufficiently large overall gate fidelity. The discussion of ground-state collisions is deferred to later in this section.

It follows from the tensor form of the electric dipole-dipole interaction that the atomic ground-state magnetic quantum numbers are not conserved, as seen in the frame transformation Eq. (17). Only in the limit of point dipoles does the BF axis coincide with the SF axis where the light shift interaction for a $\pi$-polarized laser either conserves or exchanges the quantum numbers $m_{Fi}$. The issue of minimizing leakage is discussed in detail in [8]. As shown in Eq. (14), the allowed transitions must conserve the total magnetic quantum number. Off-diagonal transitions that change the individual $m_F$ but preserve $M_{tot}$ are suppressed by the state dependent nature of the optical lattice trap. For deep wells, the potentials near the minima are approximately harmonic, and the spatial overlaps between ground-state wave functions of the different spinor components, $|\psi^{F,M_F}\rangle$, exponentially decrease with wavepacket separation (for the detailed form, see [21]). Because the dipole-dipole interaction conserves total $M_{tot}$ any transition by one atom must be accompanied by a corresponding transition in the other, e.g. from the logical $|11\rangle$ state $|\psi_0^{2,1}\rangle \otimes |\psi_0^{2,-1}\rangle \to |\psi_0^{2,2}\rangle \otimes |\psi_0^{2,-2}\rangle$, where the subscript denotes the vibrational quantum number. Thus the off-diagonal coupling is suppressed by a factor $\langle\psi_0^{2,2}|\psi_0^{2,1}\rangle\langle\psi_0^{2,-2}|\psi_0^{2,-1}\rangle = |\langle\psi_0^{2,2}|\psi_0^{2,1}\rangle|^2$. Figure 4b shows graphs of the spatial overlaps between the common external wavefunction for the logical basis states of each species atom and neighboring external states. The worst case wavefunction overlap is $|\langle\psi_0^{2,2}|\psi_0^{2,1}\rangle|^2$ and is negligible ($< 0.1$) for separations $k\Delta z > 0.38$, which at the localization $\eta = 0.05$ corresponds to separations $\Delta z > 7.6 z_0$. An additional barrier to leakage is the energy gap between ground vibrational states of different internal states. As seen in Fig. 4a, there is an effective longitudinal magnetic field due to the optical lattice itself [20]. Provided the energy uncertainty of the dipole-dipole interaction is much less than the energy gap $\Delta E$, or $\hbar/\tau \ll \Delta E$, where $\tau$ is the gate time, transitions to neighboring ground vibrational states are off-resonance. There can be appreciable coupling between initial ground-states and the excited vibrational states of neighboring wells at separations where the two energies are degenerate. An example of such a degeneracy occurs for a localization $\eta = 0.05$ and a well separation of $k\Delta z = 0.117$. In this case, the overlap amplitude between the ground motional state $|\psi_0^{F_\downarrow,1}\rangle$ and the nearly degenerate first excited motional state is $|\langle\psi_0^{F_\downarrow,1}|\psi_1^{F_\uparrow,-2}\rangle|^2 \cong 0.37$. It is thus necessary to sufficiently separate the atoms such that these leakage channels are minimized while maintaining large differential level shifts on the logical basis states.

The effect of off-diagonal leakage on the fidelity is shown in Fig. 5, which shows a sharp drop at small separations. This plot also shows the extreme sensitivity of the fidelity to atomic localization. As the atoms are more tightly trapped, the wavepackets can be brought closer together before significant overlap with Condon points occurs. For a localization $\eta = 0.05$, and at the detuning $\Delta = 10^4 \Gamma$, the peak fidelity is $\mathcal{F} = 0.925$ at $k\Delta z = 0.15$. At the same detuning but at the localization $\eta = 0.01$, the peak fidelity is $\mathcal{F} = 0.987$ at $k\Delta z = 0.078$. Such an improvement comes at the cost of increased laser trapping power as the localization scales weakly with the reciprocal of the trapping intensity, $\eta \sim I_{trp}^{-1/4}$.

In addition to photon scattering and coherent off-diagonal leakage induced by the catalysis, there are various ground-state collisional processes that can further reduce the fidelity. For example, elastic ground-state collisions, which are at the heart of the proposal discussed in [6], have the undesirable effect here of introducing phase decoherence and new coherent leakage channels. Inelastic collisions produce similar detrimental effects and/or can kick the atoms out of the trap altogether. These processes typically occur at internuclear separations that are much smaller than those required for our protocol. We can estimate the strength of collision rates by examining the dominant ground-state interactions between two spin $1/2$ alkali atoms. At low energies, the relevant interatomic potential can be written [22],

$$V(\mathbf{r}) = V_{se} + V_D + V_{SO}. \tag{19}$$

The spin-exchange terms originate from the Heisenberg interaction for electrons and arises when the charge overlap of the two atomic clouds begin to overlap. This occurs only for $kr \leq 0.02$ and therefore does not play a role in the current situation. The second term, $V_D$, describes magnetic dipole-dipole interaction of the electrons,

$$V_D = \frac{\mu_e^2}{r^3}\left(\bar{\sigma}_\alpha \cdot \bar{\sigma}_\beta - 3(\mathbf{r} \cdot \bar{\sigma}_\alpha)(\bar{\sigma}_\alpha \cdot \mathbf{r})\right), \tag{21}$$

where $\mu_e$ is the electron Bohr magneton. The last contribution, $V_{SO}$, is the second order spin-orbit interaction which is due to modification of ground-state spin interactions through distant excited electronic states of the molecule. This latter term also has exponential character and has its dominant character at even smaller interatomic separation then $V_{se}$.

For atoms with nuclear spin that are not necessarily prepared in spin polarized states $|F, m_F = \pm F\rangle$, the potential $V_{se}$ depends on the multiple scattering lengths associated with the hyperfine sublevels. The actual functional form of the exchange interaction for alkali atoms can be estimated using the formulas given in [23]. Perturbation theory shows this interaction is negligible in the current situation. The much weaker $V_{SO}$ plays even a less important role.

The dipolar interaction, $V_D$, has a long range but the calculation of its strength can be simplified by invoking the constraint that the optical lattice suppresses transitions to magnetic states trapped in wells separated in space and energy. In particular, if we invoke the selection rule, $\Delta M_{tot} = 0$ then,

$$V_D = -\frac{2\mu_e^2}{r^3} P_2(\cos\theta) \sigma_{\alpha z} \sigma_{\beta z}, \tag{23}$$

where $\theta$ is the angle between the internuclear vector $\mathbf{r}$ and the spin quantization axis. Using the Lande-Projection theorem, we find $\sigma_{\alpha z}\sigma_{\beta z} = g_{F\alpha} g_{F\beta} m_{F\alpha} m_{F\beta}$ where the Lande g-factors $g_F = \pm 1/F_\uparrow$ for $F_{\uparrow(\downarrow)}$. Our logical basis, Eq. (1), stores atoms in pairs of states with opposite sign g-factors or $m_F$ numbers meaning all logical states see a common shift from $V_D$. Thus, this interaction does not degrade our gate protocol.

A final source of decoherence can arise from excitation of motional degrees of freedom outside the computational basis. For positive catalysis detunings, the atoms are excited to repulsive states which can reshape the wavepackets over the time of the gate and then couple to higher trap vibrational states in the ground-electronic manifold. As discussed in Sec. III, these effects are highly suppressed because of the energy gap provided by the trapped vibrational levels. Corrections to this model would require us to numerically integrate the time dependent evolution of the spinor wavepackets for the two atoms in three dimensions – a nontrival task. If corrections are substantial for a specific geometry, it may be possible to design a gate that would be tolerant to motional excitation without the introduction of phase decoherence. For instance, in the context of the ion trap, quantum gates acting between two ions and a common vibrational bus mode generally entangle motional and internal degrees of freedom during the interaction. Mølmer *et al*. [24] have shown that they can be disentangled at the end of the gate by waiting the appropriate recurrence time for the harmonic oscillator states leaving only entanglement between internal states of the constituent ions.

## V. SUMMARY

We have presented a realistic protocol for implementing quantum logic with laser trapped neutral alkalis using electric dipole-dipole interactions. Both diagonal and exchange interactions can be designed to create entanglement between internal degrees of freedom. Including the hyperfine molecular structure of interacting alkalis, it is shown that the universal CPHASE gate can be executed with high fidelity given the constraints on the system such as localization and losses from photon scattering, leakage, and collisions. The specific trapping system of the optical lattice offers flexibility in terms of designing atomic wavepackets with adjustable interatomic separations, and the introduction of a catalysis laser allows the creation of "on demand" entanglement of the atoms.

Much of this research falls under the realm of molecular coherent control and in particular demonstrates the use of laser trapped atoms to probe dimer dynamics. The ability to move pairs of tightly bound atomic wavepackets together, interact the atoms, and measure the output state, can be an important diagnostic tool to study the effects of ground-ground and ground-excited state collisions. In particular, the ability to use the geometry of a trapping potential such as the optical lattice to constrain coherent leakage outside a well defined logical basis demonstrates the ability to study and control molecular interactions.

**Acknowledgements**

We gratefully acknowledge discussions with John Grondalski, Shohini Ghose, Poul Jessen and Paul Julienne. IHD and GKB acknowledge support from the National Science Foundation (Grant No. PHY-9732456) and the Office of Naval Research (Grant No. N00014-00-1-0575). This work was supported in part by the National Security Agency (NSA) and Advanced Research Development Activity (ARDA) under Army Research Office (ARO) contract number DAAD19-01-1-0648. CJW acknowledges the support of the Office of Naval Research.

# FIGURE CAPTIONS

**FIG. 1.** Two three-level atoms excited by a catalysis pulse at frequency $\omega_c$. (a) Separated noninteracting atoms. (b) Molecular eigenstates on dipole-dipole coupling. The detunings catalysis from molecular resonance at a fixed internuclear separation are indicated.

**FIG. 2.** Molecular potentials of the D1 line of $^{87}$Rb. For large $r$, the states asymptote to uncoupled atomic states, and for small $r$, to the Hund's case (c) states, as shown. Logical $|0\rangle$ and $|1\rangle$, encoded in the internal states $S_{1/2}(F=1)$ and $S_{1/2}(F=2)$ respectively, are excited by a catalysis laser, blue detuned from the transition $S_{1/2}(F=1) \to P_{1/2}(F=1)$. The differential light-shift on the logical states leads to the CPHASE gate. The relative coordinate probabality distribution is shown for two atomic Gaussian wavepackets of rms width $z_0 = 0.05\lambda$, separated by $\Delta z = 5.2 z_0$. By keeping the packets separated, resonant excitation at the Condon radius is strongly suppressed.

**FIG. 3.** Calculated fidelity, including loss from photon scattering, for a CPHASE gate via laser catalysed interaction. Fidelity $\mathcal{F}$ is plotted as a function of wavepacket separation $\Delta z$ in units of rms width $\eta = k z_0 = 0.05$, and laser detuning in units of atomic natural linewidth $\Gamma$.

**FIG. 4.** State dependent trapping in an optical lattice and suppression of leakage. (a) Trapping potentials for the localization $\eta = 0.05$. The left and right displaced solid lines correspond to trapping for the states $(F_{\downarrow,\uparrow}, m_F = \pm 1)$ and $(F_{\downarrow,\uparrow}, m_F = \mp 1)$ respectively, the long dashed lines for states $(F_\uparrow, m_F = 2)$ and $(F_\uparrow, m_F = -2)$, and the short dashed line for the states $(F_{\downarrow,\uparrow}, m_F = 0)$. (b) The wavefunction overlap between ground vibrational states of different internal states falls off exponentially with well separation $\Delta z$. The short dashed line shows the overlaps $\langle \psi_0^{F',0} | \psi_0^{F_\uparrow,1} \rangle$, the solid line $\langle \psi_0^{F_\downarrow,1} | \psi_0^{F_\uparrow,1} \rangle$, and the long dashed line $\langle \psi_0^{F_\uparrow,2} | \psi_0^{F_\uparrow,1} \rangle$. The vertical dashed line indicates the separation $k\Delta z = 0.38$ at which the largest overlap is $\left| \langle \psi_0^{F_\uparrow,2} | \psi_0^{F_\uparrow,1} \rangle \right|^2 = 0.1$.

**FIG. 5.** Calculated fidelity, including loss from photon scattering and leakage, for a CPHASE gate. The plots show fidelity at the laser detuning $\Delta = 10^4 \Gamma$ for the indicated localizations as a function of wavepacket separation in units of rms width for each localization. For comparison, the dashed line shows the calculated fidelity at $\eta = 0.05$ when leakage is not included in the model.

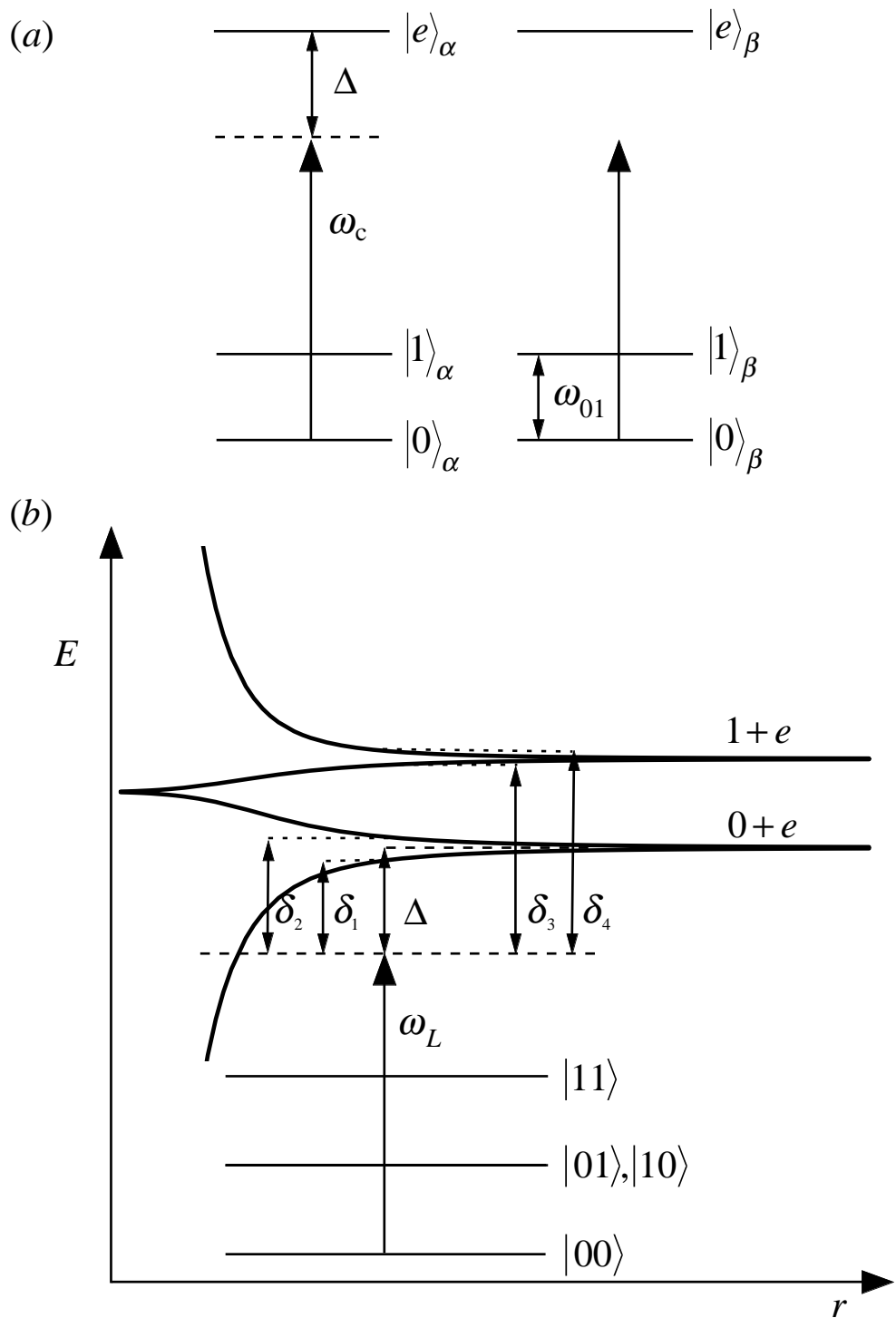

Fig.1

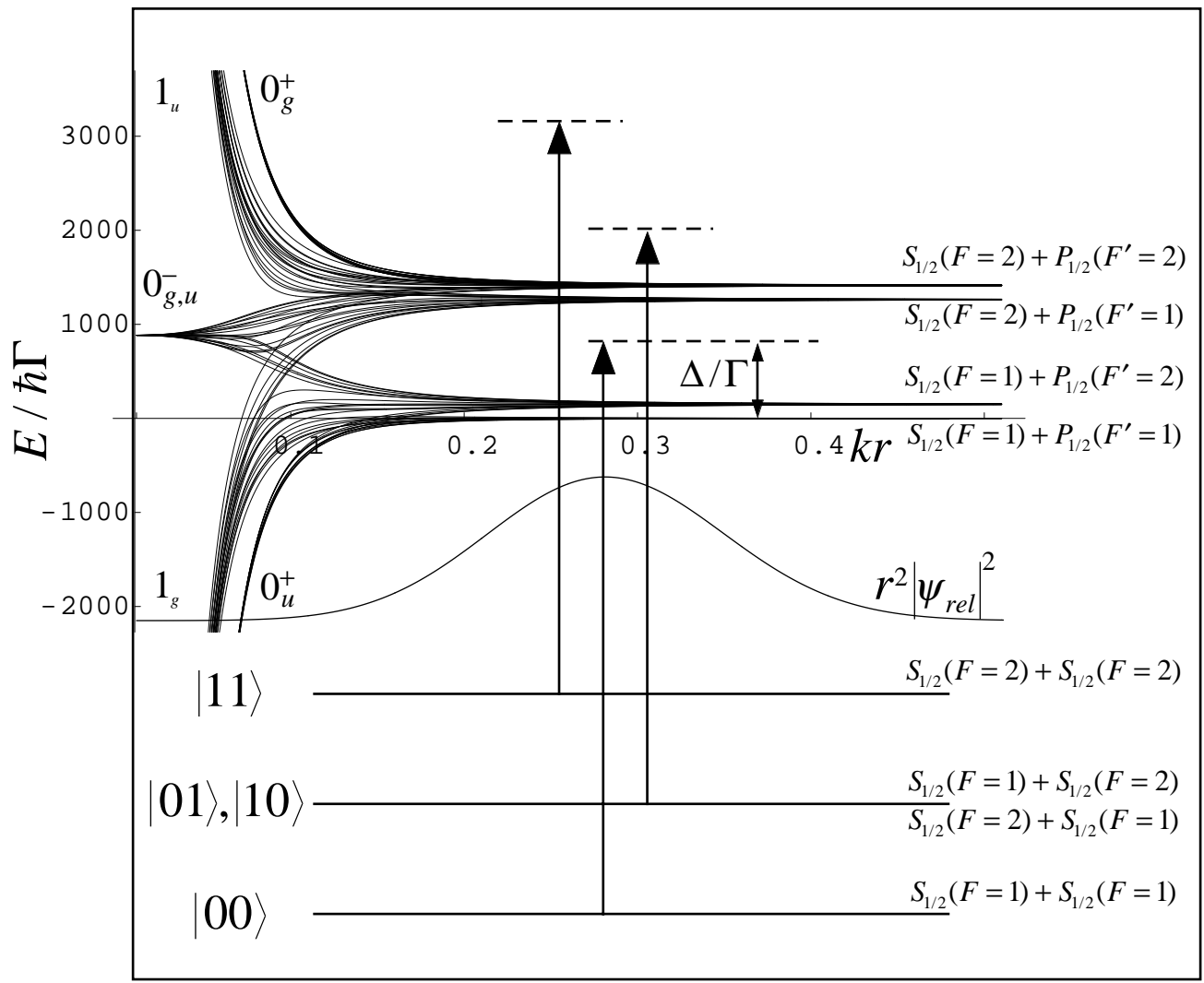

Fig.2

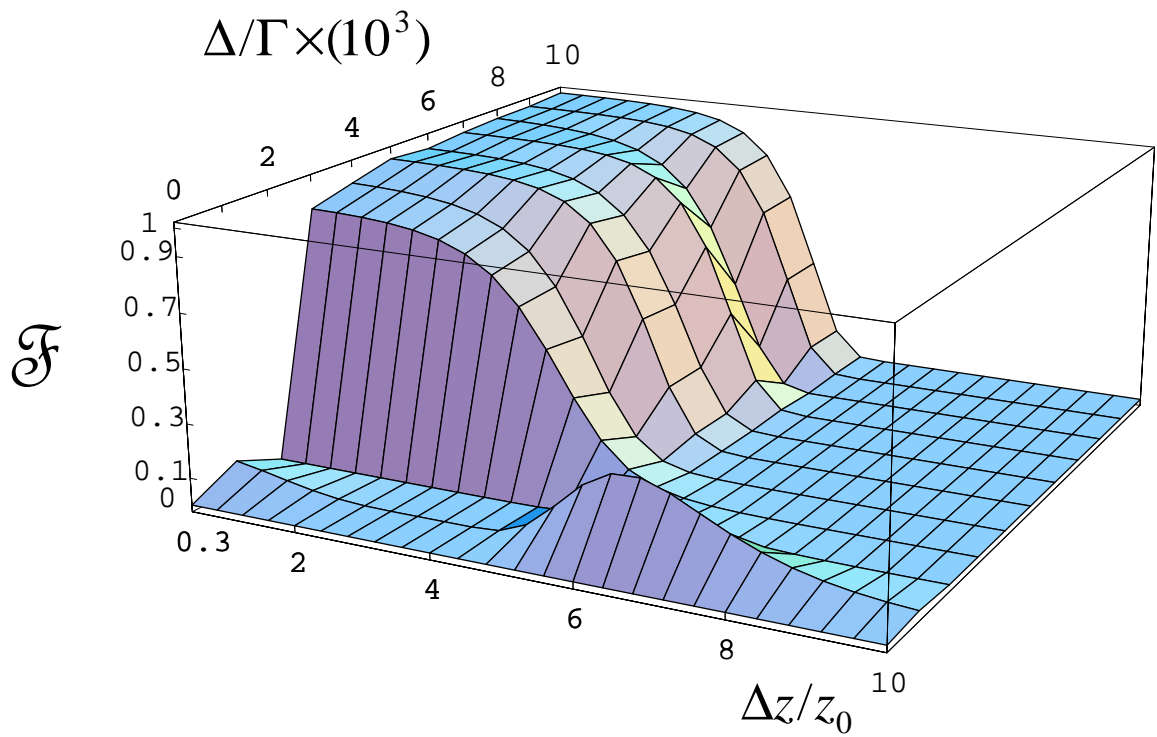

Fig.3

(a)

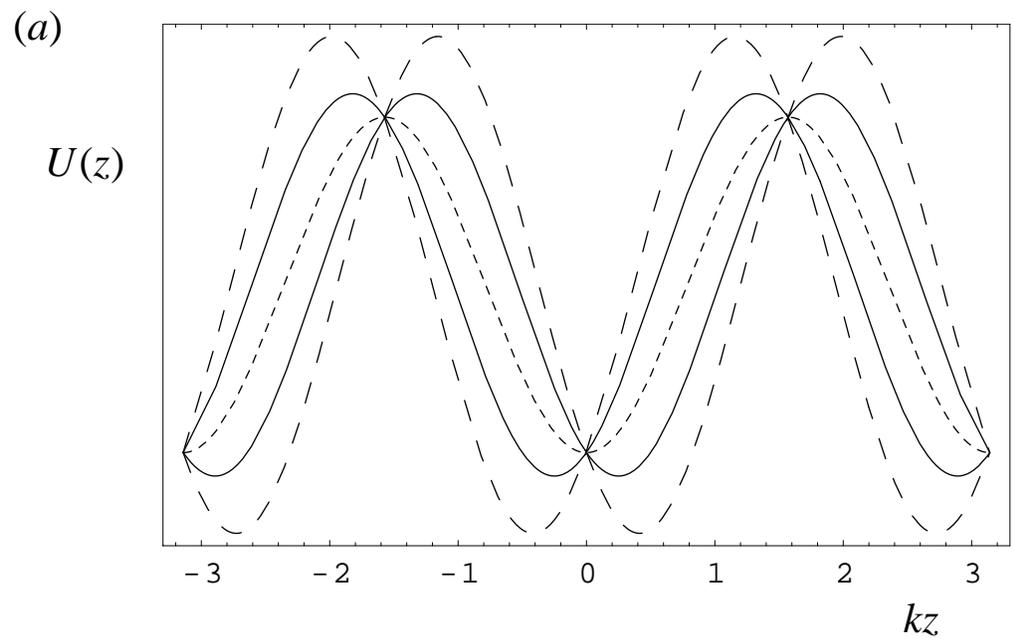

(b)

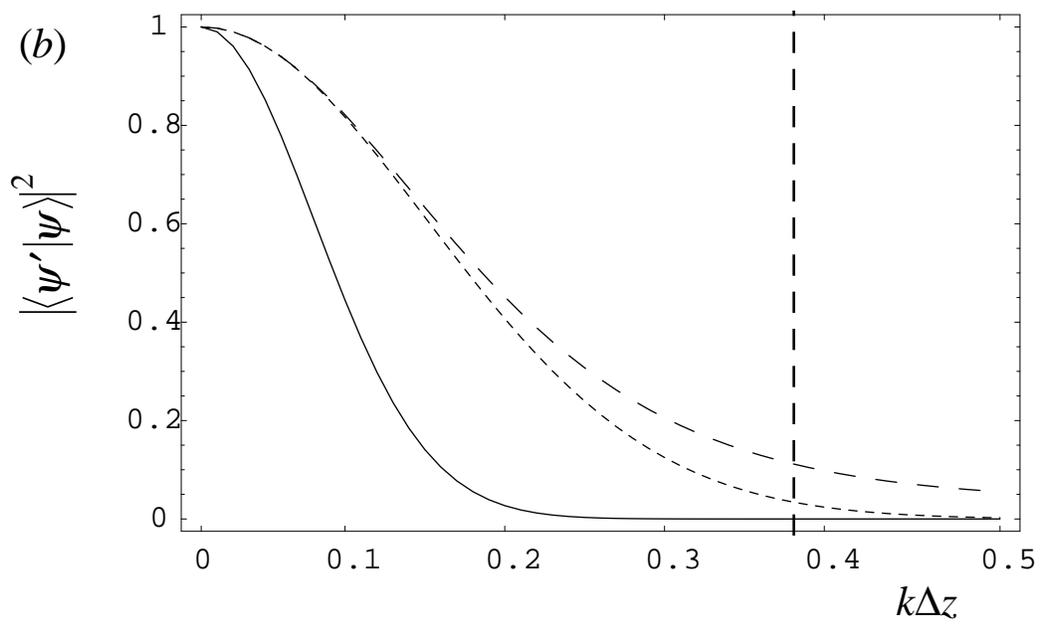

Fig.4

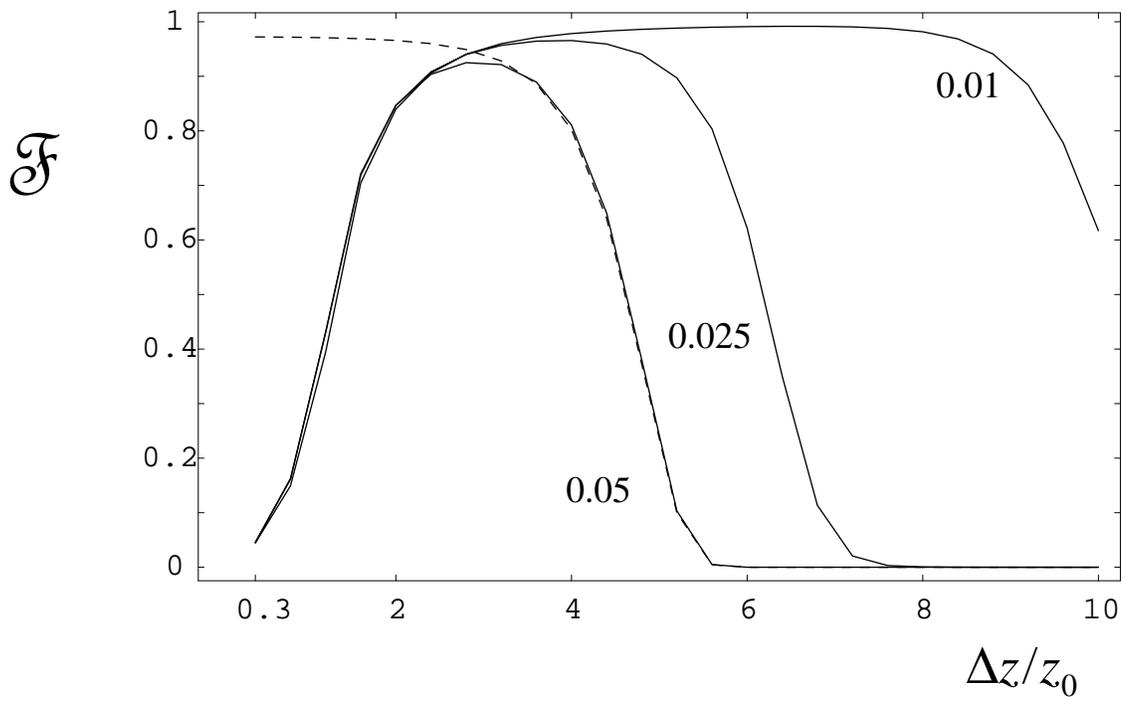

Fig.5